\documentclass[a4paper,12pt]{article}

\usepackage[utf8]{inputenc}
\usepackage[T1]{fontenc}
\usepackage[english]{babel}
\usepackage{graphicx}	
\usepackage{supertabular}
\usepackage{epsfig}
\usepackage{lscape}
\usepackage{amsmath, amsfonts}
\usepackage{mathtext}
\usepackage{times}

\begin{document}
\begin{center}
\begin{LARGE}
\vspace{3cm}
\textbf{On X-ray emission of radio pulsars \\ }
\end{LARGE} 
\vspace{1cm}
Igor F. Malov\thanks{E-mail: malov@prao.ru}, M. A. Timirkeeva \\ \textit{P.N. Lebedev Physical Institute of the Russian Academy of Sciences, Leninskii pr. 53, Moscow, 119991, Russia}
\end{center}

\begin{abstract}
The analysis of distributions of some parameters  of radio pulsars emitting X-ray radiation was carried out. The majority of such pulsars has short  spin periods  with the average value $< P >$ = 133 msec. The distribution of period derivatives reveals  a bimodality, dividing millisecond ($< log \dfrac{dP}{dt}>$ = -19.69) and normal ($< log \dfrac{dP}{dt}> $ = -13.29) pulsars. Magnetic fields at the surface of the neutron star are characterized by the bimodal distribution as well. The mean values of $<log B_s>$ are  $8.48$ and $12.41$ for millisecond pulsars and normal ones, respectively. The distribution of magnetic fields near the light cylinder, it does not show the noticeable  bimodality. The median value of $log B_{lc}$ = 4.43 is almost three orders higher comparing with this quantity ($<log B_{lc}>$ = 1.75) for radio pulsars without registered X-ray emission. Losses of rotational energy ($<log \dfrac{dE}{dt}>$ = 35.24) are also three orders higher than corresponding values for normal pulsars. There is the strong correlation between X-ray luminosities and losses of rotational energies.  The dependence of the X-ray luminosity on the magnetic field at the light cylinder has been detected. It shows that the generation of the non-thermal X-ray emission takes place at the periphery of the magnetosphere and is caused by the synchrotron mechanism. We detected the positive correlations between luminosities in radio, X-ray and gamma -ray ranges. Such correlations give the possibility to carry out a purposeful search for pulsars in one of these ranges if they radiate in  other one. 

\end{abstract}

\section{Introduction}
One of the problems in pulsar investigations remains the understanding of  the nature of  their  X-ray emission. There were some attempts to describe possible sources of this emission in pulsar magnetospheres (see, for example, \cite{Wang1998} and \cite{ZH2000}).

However, an adequate description is absent up to now.  At present there are detailed data for 61 radio pulsars from ATNF catalogue (ver. 1.58, \cite{Manchester2005}) emitting X-rays as was shown by \cite{Possenti2002, PB2015}. The X-ray observations by ROSAT (0.1-2 keV), ASCA (0.4-10 keV), XMM-Newton (0.2-12 keV) and  Chandra (0.1-10 keV) have broadened significantly our understanding of origin and mechanism generation of non-thermal and thermal radiation from a neutron star, but not completely. We propose one of the ways in advancing the solution of this problem, namely the comparison of the parameters of radio pulsars, as loud and quiet X-ray sources, revealation of their essential differences and analysis of reasons causing such differences. The main aims of our work are comparing of the known pulsar parameters of two mentioned groups, search for  their differences and analysis of possible reasons of such differences. 

The remainder of this paper is organised as follows. The used sample is presented in Section \ref{sect:sample}, also in this Section we consider some distributions of main parameters of loud and quiet pulsars. Section \ref{sect:relat} contains the analysis and the interpretation of some correlations between parameters of pulsars considered. In  Section \ref{sect:conclusion} we discuss the obtained results and give the conclusions.

\section{The used sample of pulsars and distributions of their main parameters}
\label{sect:sample}

Table \ref{Tab: Tab1} contains data for radio pulsars detected as X-ray emitters. We exclude from the consideration  anomalous X-ray pulsars (AXPs) and soft gamma-ray repeaters (SGRs). In Table \ref{Tab: Tab1} we give names of pulsars, their periods (in msec), derivatives of periods, radio luminosities at 1400 MHz (in mJy$\cdot$kpc$^2$), magnetic fields at the surface and near the light cylinder (in G), rate of  losses of  rotational energy (in erg/sec). All these characteristics describe the physical conditions in pulsars. 
 
Let us consider distributions of mentioned parameters for loud pulsars from Table \ref{Tab: Tab1}. They are presented in Fig. 1-5.

\begin{figure}
	\includegraphics[width=8cm, angle=0]{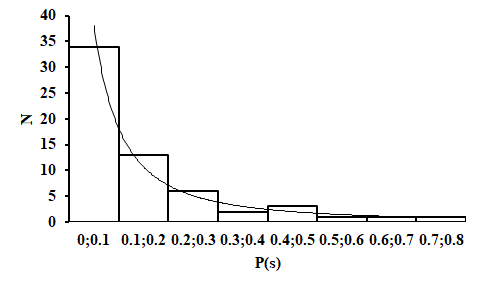}
    \caption{Distribution of pulsar spin periods for radio pulsars with the detected X-ray emission. Exponential curve is described by Eq. \ref{eq1}.}
    \label{fig:fig1}
\end{figure}

Fig. \ref{fig:fig1} shows that the majority of objects is characterized by very short periods. The mean value is $<P> = 133$ msec. The distribution of periods can be described by the following dependence:
\begin{equation} 
\label{eq1}
N(P) = ( 82 \pm 6 )  exp \left\lbrace  - P \cdot (8.92 \pm 0.60) \right\rbrace 
\end{equation}

The corresponding value of CHI-square is 1.19.

\begin{figure}
	\includegraphics[width=8cm, angle=0]{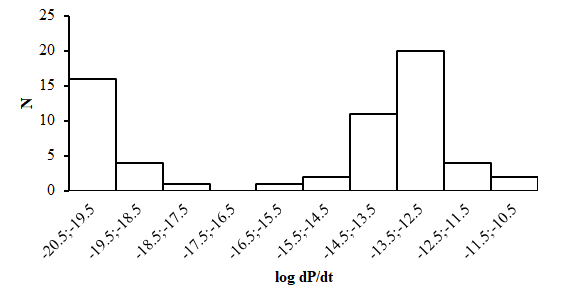}
    \caption{Distrbution of derivatives of spin periods.}
    \label{fig:fig2}
\end{figure}

The distribution of period derivatives is shown in Fig. \ref{fig:fig2}. It is bimodal. For 21 sources the mean value $< log \dfrac{dP}{dt} >$ = -19.69. They form the group of millisecond recycled pulsars. Then we can see the marked gap. For the right group with 40 objects the mean value is $< log \dfrac{dP}{dt} >$ = - 13.29. These pulsars have evolved probably as isolated objects outside any binary systems.

\begin{figure}
	\includegraphics[width=8cm, angle=0]{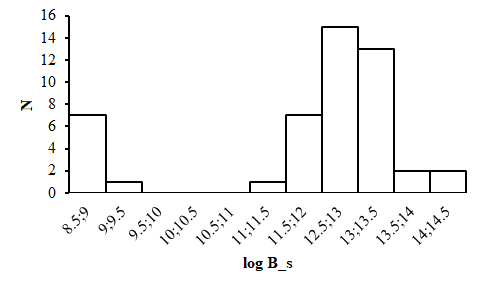}
    \caption{Distribution of magnetic fields at the surface.}
    \label{fig:fig3}
\end{figure}

For an isolated (non-accreting) pulsar the magnetic dipole field strength can be estimated from the observed period P, and its period derivative $ \dfrac{dP}{dt}$, as 
\begin{equation} 
B_s \simeq 3.2 \times 10^{19} \sqrt{P \dfrac{dP}{dt}}
\end{equation}

For accreting pulsars such an estimate can be erroneous, because their rotation can be accelerated as a result of the transfer of the angular momentum from the falling matter.

The distribution of the magnetic fields at the surface looks similarly to the distribution of their period derivatives (Fig. \ref{fig:fig3}). It is bimodal. The right part in Fig.  \ref{fig:fig3} contains normal pulsars with $< log B_s >$ = 12.43. The left part presents the population of millisecond recycled pulsars with $< log B_s >$ = 8.46. Evolution of such pulsars had been described by \cite{bkk1974}. They showed that there was the relationship between the final spin period and the corresponding magnetic field:
\begin{equation} 
\label{eq2}
P_0 =2.6  B_{12}^{6/7}  L_{36}^{-3/7}
\end{equation}

In Eq. \ref{eq2} the standard denotations are used, such as $B_{12}$ is the magnetic field strength in units of 10$^{12}$ G and $L_{36}$ is the X-ray luminosity of this system in units of 10$^{36}$ erg/sec.

\begin{figure}
	\includegraphics[width=8cm, angle=0]{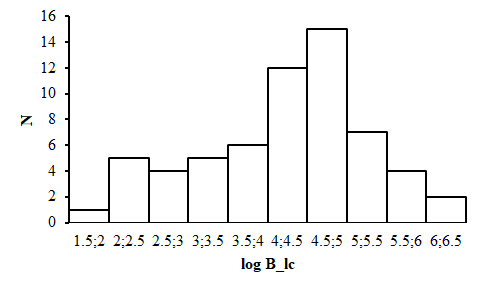}
    \caption{Distribution of magnetic fields at the light cylinder. }
    \label{fig:fig4}
\end{figure}

In Fig. \ref{fig:fig4} the distribution of magnetic fields at the light cylinder is shown. It is suggested that the magnetic field is dipolar from the surface of the neutron star up to the periphery of the magnetosphere:

\begin{equation} 
\label{eq3}
B = B_s \Big(\dfrac{R_*}{r_{lc}}\Big)^3 \text{,}
\end{equation}

where R$_{*}$ is the neutron star radius,
 
\begin{equation} 
\label{eq4}
r_{lc}=\dfrac{cP}{2\pi} 
\end{equation}

is the radius of the light cylinder. This distribution can be described by the gaussian with the maximum at $< log B_{lc} >$ = 4.84, the width equaled to 0.90 and the median value $ log B_{lc} $ = 4.43. The CHI-square for this inscribing is 2.27. For normal quiet pulsars values of magnetic fields at the light cylinder are almost three orders less than the obtained median value \cite{mt2014}.

\begin{figure}
	\includegraphics[width=8cm, angle=0]{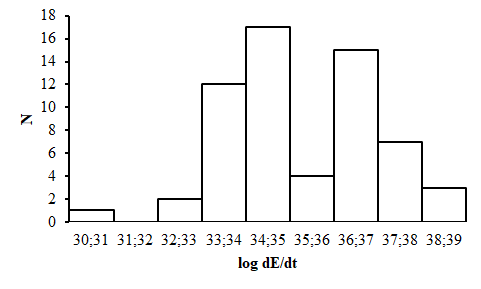}
    \caption{Distribution of rates of losses of rotational energy for the considered sample.}
    \label{fig:fig5}
\end{figure}

Fig. \ref{fig:fig5} shows the distribution of losses of rotational energies

\begin{equation}
\frac{dE}{dt} = \frac{4  \pi^2 I \frac{dP}{dt}}{P^3} \text{,}
\label{eq5}
\end{equation}

where I is the moment of inertia of the neutron star. This distribution is bimodal also. It is the consequence of the bimodality of the distribution for $\dfrac{dP}{dt}$. It is worth noting that these values of $\dfrac{dE}{dt}$ are as a rule much higher than corresponding values for quiet pulsars. The mean value for objects of Table \ref{Tab: Tab1} is $<\dfrac{dE}{dt}>$ = 35.24, the median value is 34.72.

We showed earlier \cite{mt2014,mtasp2017} that rates of losses of rotational energy for gamma pulsars on average had been also much higher than for gamma quiet radio pulsars ($<\dfrac{dE}{dt}>$ = 35.53 and 32.60, respectively). Their magnetic fields at the light cylinder are strongly differ as well. The corresponding values are $<B_{lc}$ > = 9 $\times$ 10$^3$ G and 56 G. This means that hard radiation is registered as a rule in pulsars with high values of $\dfrac{dE}{dt}$ and B$_{lc}$. The search for X-ray and gamma-ray emission in future can be carried out purposefully using for this aim radio pulsars with the mentioned peculiarities (see also \cite{mt2018}).

\section{Analysis of some relationships between parameters of pulsars considered}
\label{sect:relat}

All these objects have been observed on boards of X-ray space satellites. It allows us to add their estimates of luminosity \cite{Possenti2002,PB2015} in our sample of pulsars. These authors have used two energy bands (0.1-2 KeV and 2-10 keV) for X-ray pulsars. As for the pulsar J2022+3842, its X-ray flux density (\textit{F}) was taken from the other paper \cite{apk2014}, and the X-ray luminosity has been calculated by multiplying \textit{F} by $d^2$ (the distance is taken from \cite{Manchester2005}).  

\begin{figure}
	\includegraphics[width=8cm, angle=0]{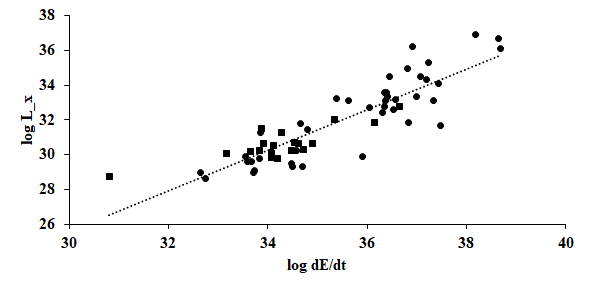}
    \caption{Relationship between X-ray luminosity and the rate of losses of rotational energy.}
    \label{fig:fig6}
\end{figure}

First of all let us consider the dependence of the X-ray luminosity on the rate of losses of rotational energy (Fig. \ref{fig:fig6}). For our sample this dependence can be described by the following equation:

\begin{equation}
\label{eq6}
log L_x = (1.17 \pm 0.08) log \dfrac{dE}{dt} -9.46 \pm 2.89 \text{,}
\end{equation}

the correlation coefficient is $K = 0.97$, and the probability of the random distribution is $p < 10^{-4}$.

Constructing Fig. \ref{fig:fig6}, we have used values of luminosities in the range 2-10 keV \cite{Possenti2002} (dots) and 0.1-2 keV \cite{PB2015} (squares). All the X-ray emission in the range 2-10 keV is non-thermal. For the diapason from 0.1 to 2 keV we have taken only the part of emission described by the power-law. It is caused by non-thermal mechanisms as well. As we can see in Fig. \ref{fig:fig6}, both ranges are well described by the unique dependence:

\begin{equation}\label{eq7}
L_x = 3.47 \times 10^{-10} \Big(\dfrac{dE}{dt}\Big)^{1.17}
\end{equation}

The relationship $L_x (\frac{dE}{dt})$ has been analysed many times in a number of previous papers for different ranges and different samples. For example, \cite{Possenti2002} obtained the following equation:

\begin{equation} \label{eq8}
log L_{x(2-10)} = 1.34  log \frac{dE}{dt} - 15.30
\end{equation}

and mentioned two other results, obtained by \cite{ZH2000}:
 
\begin{equation} \label{eq9}
L_{x (0.1-2.4)} = 10^{-3}  \Big(\frac{dE}{dt}\Big)  
\end{equation}

and
 
\begin{equation} \label{eq10}
L_{x(2-10)} = 10^{-21} \Big(\frac{dE}{dt} \Big)^{\frac{3}{2}} 
\end{equation}

Thermal X-ray emission, detected in some pulsars \cite{PB2015} can be connected with
a hot surface of the neutron star (T  $\sim~ 10^6$ K), with the heating of a polar cap due to bombarding it by accelerated positrons generated in a cascade process of gamma quanta conversion (T$\sim~ 10^7$ K), with an accretion from a relic disk or a disk raked up during the movement of the neutron star through the interstellar medium. Partially such radiation can be connected with the supernova remnant.

The non-thermal radiation can be explained by the appearance of noticeable pitch-angles of relativistic electrons at the periphery of the magnetosphere near the light cylinder and by the switching on the synchrotron mechanism \cite{MM2002}. In this case the X-ray luminosity can be calculated using the following formula:

\begin{equation} \label{eq11}
L_{x} \approx \frac{16 \pi^8 e^4 R_*^6 I B_s^2 \gamma_r \Psi^2 \frac{dP}{dt} sin ^4\beta}{m^3 c^{11} P^8 } \text{,}
\end{equation}

where $\gamma_r$ is the Lorenz-factor of emitting particles, $\Psi$ is their pitch-angle, $\beta$ is the angle between the magnetic moment and the rotation axis. The magnetic moment is believed now as the axis of the emission cone. \cite{MM2002} obtained the following expression:

\begin{equation} \label{eq12}
\Psi = \dfrac{1}{2} \Big( \frac{3 \pi^3 m^5 c^7 \gamma_b ^3}{4 e^6 B^{4} P^3 \gamma_p ^4  \gamma_r ^2 } \Big) 
\end{equation}

We can calculate the expected values of X-ray luminosities using the following formula:

\begin{equation} \label{eq13}
L_{calc} = \dfrac{\sqrt{3} \pi^{7/2} e I \frac{dP}{dt} \gamma_b ^{3/2} }{32 c^{3/2} P^{7/2} m^{1/2} \gamma_p^{2} } = 10^{27} \frac{ \Big(\frac{dP}{dt}\Big)_{-15} }{ P^{7/2} } \text{erg/sec} 
\end{equation}

\begin{figure}
	\includegraphics[width=8cm, angle=0]{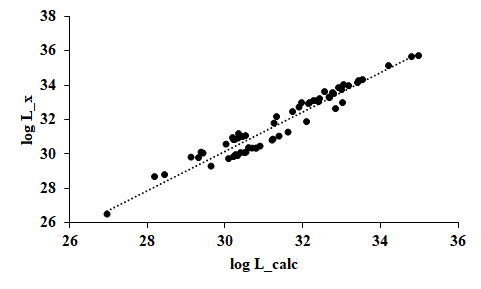}
    \caption{Comparison of observed and model values of X-ray luminosities.}
    \label{fig:fig7}
\end{figure}

The comparison of calculated and observed luminosities is given in Fig. \ref{fig:fig7}. In equations \ref{eq12} and \ref{eq13} $\gamma_b$ is the Lorenz-factor of the primary beam, $\gamma_p$ is the Lorenz-factor of the secondary electrons. We suggest that $\gamma_b = 5 \times 10^6$, $\gamma_p$ = 10 for all pulsars considered.
The line in Fig. \ref{fig:fig7} corresponds to the relationship

\begin{equation} \label{eq14}
L_x = 3.48 \times 10^{-4} L_{calc} ^{1.14} 
\end{equation}
 or

\begin{equation} \label{eq15}
log L_x = (1.14 \pm 0.04) log L_{calc} - 3.46 \pm 1.20 \text{,}
\end{equation}

with the correlation coefficient $K = 0.97$. Taking into account that values of $\gamma_b$ and $\gamma_p$ in different pulsars can differ the agreement between $L_{calc}$ and $L_x$ must be recognized as very good, and the used synchrotron model as adequately describing the data of observations.

\begin{figure}
	\includegraphics[width=8cm, angle=0]{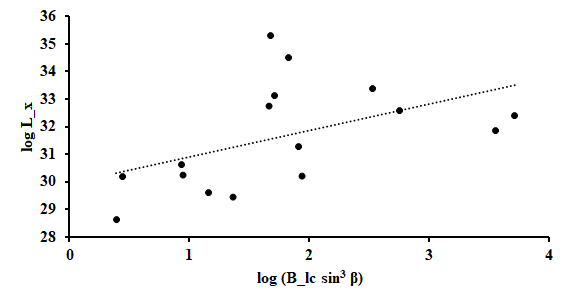}
    \caption{Dependence of X-ray luminosity on magnetic field at the periphery magnetosphere.}
    \label{fig:fig8}
\end{figure}

\begin{figure}
	\includegraphics[width=8cm, angle=0]{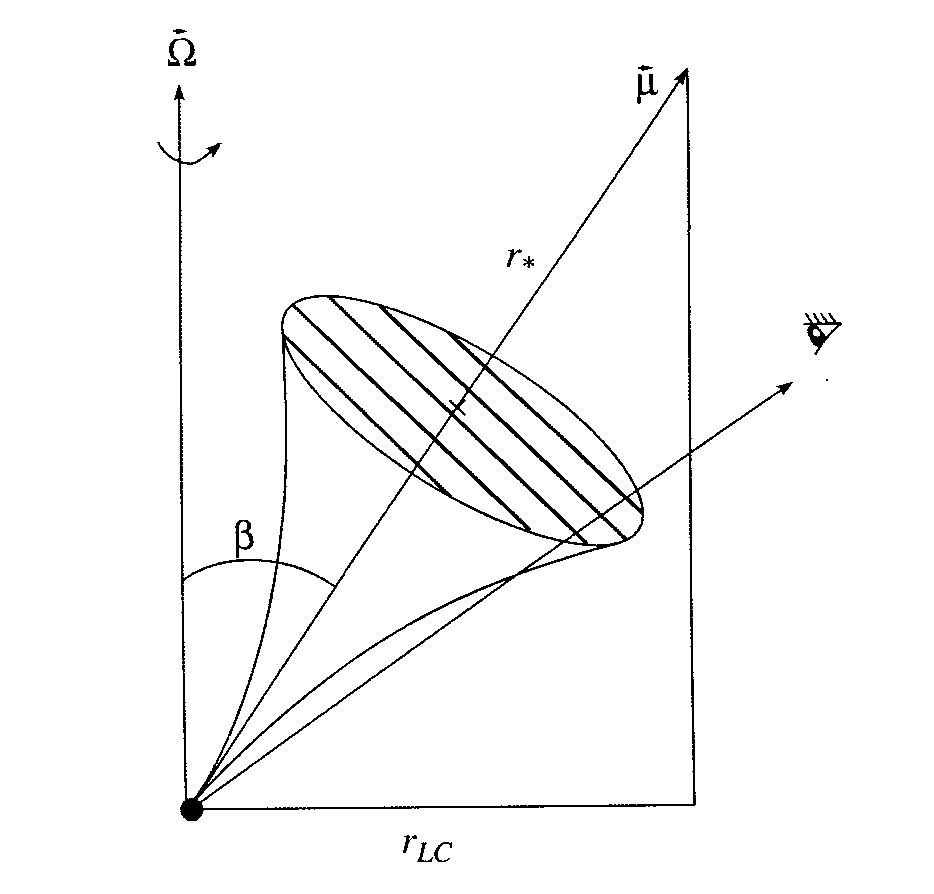}
    \caption{Model of the magnetosphere.}
    \label{fig:fig9}
\end{figure}

Another argument for the conclusion on the generation of non-thermal X-ray emission at the periphery of the pulsar magnetosphere by the synchrotron mechanism is the noticeable correlation between the X-ray luminosity and the magnetic field near the light cylinder (Fig. \ref{fig:fig8}). Here we have taken into account that the scale of the magnetosphere can differ from the radius of the light cylinder, and the distance where the observed radiation is generated depends on the angle $\beta$ (Fig. \ref{fig:fig9}):

\begin{equation}
\label{eq16}
R_{*} = \dfrac{r_{lc}}{sin \beta}
\end{equation}

In pulsars with a small angle $\beta$ the region of the formation of the observed radiation extends much further than in the pulsars with a large inclination of the axes. We added estimates of angles $\beta$ (in degrees) from \cite{NM2017} in penultimate column of Table \ref{Tab: Tab1}. 

The corresponding relationship in Fig. \ref{fig:fig9} can be presented as

\begin{equation}
\label{eq17}
log L_x = (0.96 \pm 0.46) log (B_{lc} sin^3 \beta ) + 29.93 \pm 0.92 
\end{equation}

The correlation coefficient is $K = 0.49$ and the probability of the random distribution is $p \sim 5.5 \%$.

\begin{figure}
	\includegraphics[width=8cm, angle=0]{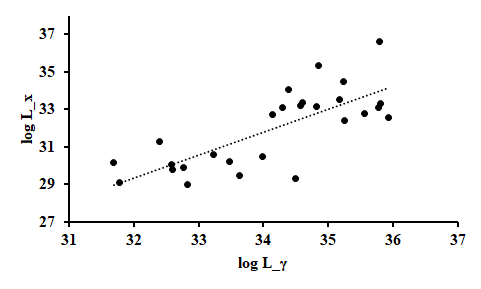}
    \caption{Comparison of X-ray and gamma-ray luminosities.}
    \label{fig:fig10}
\end{figure}

We have shown early \cite{mt2015} that there is the similar correlation between gamma-ray luminosities and magnetic fields at the light cylinder. Hence, we must expect the noticeable correlation between gamma-ray and X-ray luminosities. To compare these quantities we used data from the catalogue of \cite{Abdo2013}. Fig. \ref{fig:fig10} shows that indeed there is strong correlation between L$_x$ and L$_\gamma$:

\begin{equation}
\label{eq18}
log L_x = (1.22 \pm 0.21) log L_{\gamma} - 9.67 \pm 7.18 
\end{equation}

The correlation coefficient is $K = 0.76$ and the probability of the random distribution is $p < 10^{-4}$.

\begin{figure}
	\includegraphics[width=8cm, angle=0]{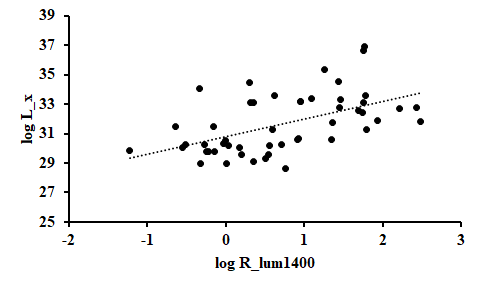}
    \caption{Comparison of radio and X-ray luminosities.}
    \label{fig:fig11}
\end{figure}

The comparison of radio and X-ray luminosities is presented in Fig. \ref{fig:fig11}. We can see that these quantities correlate as well:

\begin{equation}
\label{eq19}
log L_x = (1.20 \pm 0.28) log R_{lum1400} +30.80 \pm 0.31
\end{equation}
The correlation coefficient is $K = 0.53$.
The obtained correlations can be used for a purposeful search for radiation in one of the range from sources detected in another range \cite{mt2015,mt2018}.

\section{Conclusions and discussion}
\label{sect:conclusion}
\begin{itemize}
\item The majority of radio pulsars with the registered X-ray emission has short spin periods with $<P>$ = 133 msec.
\item The distribution of period derivatives shows the bimodality. One group of objects contains millisecond recycled pulsars with $<log\dfrac{dP}{dt}>$= -19.69, another one includes normal pulsars with $<log\dfrac{dP}{dt}>$ = -13.29.
\item The similar bimodality is seen in the distribution of magnetic fields at the surface of the neutron star. The mean values of log B$_s$ are 8.46 and 12.43 for millisecond and normal pulsars, respectively.
\item The distribution of magnetic fields at the light cylinder does not show the noticeable bimodality. Instead it can be presented by the unique gaussian. The median value of $<log B_{lc} >$ = 4.43 is almost three orders of magnitude higher than the corresponding values for radio pulsars without X-rays emission ($<log B_{lc}>$ = 1.75.
\item The rate of losses of rotational energy for pulsars considered ($<log \dfrac{dE}{dt}>$ = 35.24) is also three orders higher than the corresponding values for pulsar quiet in X-rays.
\item As was expected there was the strong correlation between X-ray luminosities of radio pulsars and the rates of losses of their rotational energy. The last ones are believed as the main source of energy for all processes in the pulsar magnetosphere.
\item The dependence of X-ray luminosities on magnetic fields at the light cylinder is detected. This dependence shows that the generation of X-ray emission takes place at the periphery of the magnetosphere and is caused by the synchrotron mechanism.
\end{itemize}
The obtained results lead to the conclusion that the division of loud pulsars into 5 groups proposed by \cite{Possenti2002} is not necessary. In fact there are two populations. The first one includes pulsars with long spin periods and weak or absent X-ray radiation. They can emit thermal radiation from the surface. The second population contains objects with rather short periods. They are characterized by high magnetic fields near the light cylinder. This leads to the switching on the synchrotron mechanism and generation of non-thermal X-ray emission. The inverse Compton scattering of soft X-ray quanta on relativistic electrons can explain gamma-ray emission up to energies of hundreds GeV and may be even TeVs. The detected correlations between luminosities in different ranges give the possibility for a purposeful search for new pulsars in all diapasons if there is the registered radiation in one of these ranges.

\section*{Acknowledgements}
This work has been carried out with the financial support of Basic Research Program of the Presidium of the Russian Academy of Sciences "Transition and Explosive Processes in Astrophysics (P-41)" and Russian Foundation for Basic Research (grant 16-02-00954).



\newpage

\begin{landscape}
\tablehead{
\multicolumn{14}{p{1.146\linewidth}}{Table 1. Considered sample of pulsars} \\
\hline
 & PSRJ    & P   & $\dfrac{dP}{dt}$ & R$_{lum1400}$ &  B$_s$  &  $\dfrac{dE}{dt}$ &  B$_{lc}$ &	log L$_x$	&	log L$_x$	&	log L$_\gamma$	&	$\beta$	&	log L$_{calc}$	\\
 &    & msec  &   & mJy $\times$ kpc$^2$ & G  &  erg/sec & G &	(2-10 keV)	& (0.1-2 keV)	&		&		&		\\ 
\hline
}
\begin{supertabular}{|c|c|c|c|c|c|c|c|c|c|c|c|c|c|}
\label{Tab: Tab1}
1	&	J0030+0451	&	4.87	&	1.02E-20	&	0.06	&	2.25E+08	&	3.5E+33	&	1.83E+04	&	29.88	&		&	32.76	&		&	30.10	\\
2	&	J0101-6422	&	2.57	&	5.16E-21	&	0.28	&	1.17E+08	&	1.2E+34	&	6.42E+04	&		&	30.04	&	32.58	&		&	30.78	\\
3	&	J0117+5914	&	101.44	&	5.85E-15	&	0.94	&	7.80E+11	&	2.2E+35	&	7.00E+03	&	30.34	&	32.04	&		&		&	31.25	\\
4	&	J0205+6449	&	65.72	&	1.94E-13	&	0.46	&	3.61E+12	&	2.7E+37	&	1.19E+05	&	34.08	&		&	34.38	&		&	33.43	\\
5	&	J0218+4232	&	2.32	&	7.74E-20	&	8.93	&	4.29E+08	&	2.4E+35	&	3.21E+05	&	33.20	&		&	34.58	&		&	32.11	\\
6	&	J0337+1715	&	2.73	&	1.77E-20	&		&	2.22E+08	&	3.4E+34	&	1.02E+05	&		&	30.71	&		&		&	31.22	\\
7	&	J0358+5413	&	156.38	&	4.39E-15	&	23.00	&	8.39E+11	&	4.5E+34	&	2.06E+03	&	31.76	&		&		&		&	30.46	\\
8	&	J0437-4715	&	5.76	&	5.73E-20	&	3.66	&	5.81E+08	&	1.2E+34	&	2.85E+04	&	30.19	&		&	31.69	&		&	30.60	\\
9	&	J0534+2200	&	33.39	&	4.21E-13	&	56.00	&	3.79E+12	&	4.5E+38	&	9.55E+05	&	36.65	&		&	35.79	&		&	34.79	\\
10	&	J0537-6910	&	16.12	&	5.18E-14	&	0.00	&	9.25E+11	&	4.9E+38	&	2.07E+06	&	36.11	&		&		&		&	34.99	\\
11	&	J0538+2817	&	143.16	&	3.67E-15	&	3.21	&	7.33E+11	&	4.9E+34	&	2.34E+03	&	29.31	&		&		&		&	30.52	\\
12	&	J0540-6919	&	50.57	&	4.79E-13	&	59.28	&	4.98E+12	&	1.5E+38	&	3.61E+05	&	36.93	&		&		&		&	34.22	\\
13	&	J0543+2329	&	246.00	&	1.54E-14	&	21.90	&	1.97E+12	&	4.1E+34	&	1.24E+03	&		&	30.61	&		&	11	&	30.32	\\
14	&	J0633+1746	&	237.10	&	1.10E-14	&		&	1.63E+12	&	3.2E+34	&	1.15E+03	&	29.33	&		&	34.50	&		&	30.23	\\
15	&	J0659+1414	&	384.89	&	5.50E-14	&	0.31	&	4.66E+12	&	3.8E+34	&	7.66E+02	&	30.26	&		&		&	13	&	30.19	\\
16	&	J0751+1807	&	3.48	&	7.79E-21	&	3.94	&	1.67E+08	&	7.3E+33	&	3.71E+04	&	31.29	&		&	32.40	&		&	30.50	\\
17	&	J0826+2637	&	530.66	&	1.71E-15	&	1.02	&	9.64E+11	&	4.5E+32	&	6.05E+01	&	28.99	&		&		&		&	28.20	\\
18	&	J0835-4510	&	89.33	&	1.25E-13	&	86.24	&	3.38E+12	&	6.9E+36	&	4.45E+04	&	31.86	&		&		&		&	32.77	\\
19	&	J0922+0638	&	430.63	&	1.37E-14	&	5.08	&	2.46E+12	&	6.8E+33	&	2.89E+02	&		&	30.23	&		&	42	&	29.42	\\
20	&	J0953+0755	&	253.07	&	2.30E-16	&	5.72	&	2.44E+11	&	5.6E+32	&	1.41E+02	&	28.62	&		&		&	15	&	28.45	\\
21	&	J1012+5307	&	5.26	&	1.71E-20	&	1.57	&	3.04E+08	&	4.7E+33	&	1.96E+04	&	29.58	&		&		&		&	30.21	\\
22	&	J1024-0719	&	5.16	&	1.86E-20	&	2.23	&	3.13E+08	&	5.3E+33	&	2.13E+04	&	29.09	&		&	31.78	&		&	30.27	\\
23	&	J1044-5737	&	139.03	&	5.46E-14	&		&	2.79E+12	&	8.0E+35	&	9.73E+03	&	29.92	&		&		&		&	31.74	\\
24	&	J1048-5832	&	123.67	&	9.63E-14	&	54.66	&	3.49E+12	&	2.0E+36	&	1.73E+04	&	32.40	&		&	35.25	&	42	&	32.16	\\
25	&	J1057-5226	&	197.11	&	5.83E-15	&		&	1.09E+12	&	3.0E+34	&	1.33E+03	&	29.48	&		&	33.63	&	15	&	30.23	\\
26	&	J1105-6107	&	63.19	&	1.58E-14	&	4.18	&	1.01E+12	&	2.5E+36	&	3.76E+04	&	33.55	&		&	35.18	&		&	32.40	\\
27	&	J1112-6103	&	64.96	&	3.15E-14	&	28.35	&	1.45E+12	&	4.5E+36	&	4.95E+04	&		&	32.78	&	35.56	&		&	32.65	\\
28	&	J1119-6127	&	407.96	&	4.02E-12	&	56.45	&	4.10E+13	&	2.3E+36	&	5.66E+03	&	33.13	&		&	35.78	&	12	&	31.97	\\
29	&	J1124-5916	&	135.48	&	7.53E-13	&	2.00	&	1.02E+13	&	1.2E+37	&	3.85E+04	&	34.48	&		&	35.23	&		&	32.92	\\
30	&	J1224-6407	&	216.48	&	4.95E-15	&	62.40	&	1.05E+12	&	1.9E+34	&	9.68E+02	&		&	31.28	&		&	26	&	30.02	\\
31	&	J1301-6310	&	663.83	&	5.64E-14	&	0.23	&	6.19E+12	&	7.6E+33	&	1.98E+02	&		&	31.48	&		&		&	29.37	\\
32	&	J1341-6220	&	193.34	&	2.53E-13	&	301.64	&	7.08E+12	&	1.4E+36	&	9.18E+03	&		&	31.85	&		&	47	&	31.90	\\
33	&	J1420-6048	&	68.18	&	8.32E-14	&	28.53	&	2.41E+12	&	1.0E+37	&	7.13E+04	&	33.33	&		&	35.81	&		&	33.00	\\
34	&	J1513-5908	&	151.25	&	1.53E-12	&	18.20	&	1.54E+13	&	1.7E+37	&	4.17E+04	&	35.32	&		&	34.85	&	6	&	33.06	\\
35	&	J1600-3053	&	3.60	&	9.50E-21	&	8.10	&	1.87E+08	&	8.1E+33	&	3.77E+04	&		&	30.61	&	33.23	&		&	30.53	\\
36	&	J1617-5055	&	69.36	&	1.35E-13	&		&	3.10E+12	&	1.6E+37	&	8.70E+04	&	34.31	&		&		&		&	33.19	\\
37	&	J1658-5324	&	2.44	&	1.12E-20	&	0.54	&	1.67E+08	&	3.0E+34	&	1.08E+05	&		&	30.23	&	33.48	&		&	31.19	\\
38	&	J1709-4429	&	102.46	&	9.30E-14	&	49.35	&	3.12E+12	&	3.4E+36	&	2.72E+04	&	32.58	&		&	35.93	&	16	&	32.43	\\
39	&	J1730-2304	&	8.12	&	2.02E-20	&	1.50	&	4.10E+08	&	1.5E+33	&	7.17E+03	&		&	30.08	&		&		&	29.62	\\
40	&	J1731-1847	&	2.34	&	2.54E-20	&	8.45	&	2.47E+08	&	7.8E+34	&	1.80E+05	&		&	30.64	&		&		&	31.61	\\
41	&	J1744-1134	&	4.07	&	8.93E-21	&	0.48	&	1.93E+08	&	5.2E+33	&	2.68E+04	&	28.97	&		&	32.83	&		&	30.32	\\
42	&	J1801-2451	&	124.92	&	1.28E-13	&	12.27	&	4.04E+12	&	2.6E+36	&	1.95E+04	&	33.37	&		&	34.60	&	15	&	32.27	\\
43	&	J1803-2137	&	133.67	&	1.34E-13	&	269.10	&	4.29E+12	&	2.2E+36	&	1.68E+04	&	32.75	&		&		&	8	&	32.19	\\
44	&	J1811-1925	&	64.67	&	4.40E-14	&		&	1.71E+12	&	6.4E+36	&	5.92E+04	&	34.93	&		&		&		&	32.81	\\
45	&	J1816+4510	&	3.19	&	4.31E-20	&		&	3.75E+08	&	5.2E+34	&	1.08E+05	&		&	30.32	&		&		&	31.37	\\
46	&	J1824-2452A	&	3.05	&	1.62E-18	&	60.50	&	2.25E+09	&	2.2E+36	&	7.40E+05	&	33.56	&		&		&		&	33.01	\\
47	&	J1825-0935	&	769.01	&	5.25E-14	&	1.08	&	6.43E+12	&	4.6E+33	&	1.33E+02	&		&	30.20	&		&	16	&	29.12	\\
48	&	J1826-1334	&	101.49	&	7.53E-14	&	27.37	&	2.80E+12	&	2.8E+36	&	2.51E+04	&	34.51	&		&		&	8	&	32.35	\\
49	&	J1832-0836	&	2.72	&	8.28E-21	&	0.72	&	1.52E+08	&	1.6E+34	&	7.08E+04	&		&	29.75	&		&		&	30.90	\\
50	&	J1846-0258	&	326.57	&	7.11E-12	&		&	4.88E+13	&	8.1E+36	&	1.31E+04	&	36.22	&		&		&		&	32.55	\\
51	&	J1856+0113	&	267.44	&	2.08E-13	&	2.07	&	7.55E+12	&	4.3E+35	&	3.70E+03	&	33.14	&		&		&		&	31.32	\\
52	&	J1911-1114	&	3.63	&	1.40E-20	&	0.57	&	2.28E+08	&	1.2E+34	&	4.48E+04	&		&	29.81	&		&		&	30.69	\\
53	&	J1932+1059	&	226.52	&	1.16E-15	&	3.46	&	5.18E+11	&	3.9E+33	&	4.18E+02	&	29.60	&		&		&	19	&	29.32	\\
54	&	J1939+2134	&	1.56	&	1.05E-19	&	161.70	&	4.09E+08	&	1.1E+36	&	1.02E+06	&	32.73	&		&	34.15	&		&	32.85	\\
55	&	J1952+3252	&	39.53	&	5.84E-15	&	9.00	&	4.86E+11	&	3.7E+36	&	7.38E+04	&	33.16	&		&	34.82	&		&	32.68	\\
56	&	J2017+0603	&	2.90	&	7.99E-21	&	0.98	&	1.54E+08	&	1.3E+34	&	5.94E+04	&		&	30.52	&	33.99	&		&	30.79	\\
57	&	J2022+3842	&	48.58	&	8.61E-14	&		&	2.07E+12	&	3.0E+37	&	1.69E+05	&	31.68	&		&		&		&	33.53	\\
58	&	J2124-3358	&	4.93	&	2.06E-20	&	0.61	&	3.22E+08	&	6.8E+33	&	2.52E+04	&	29.77	&		&	32.60	&		&	30.39	\\
59	&	J2222-0137	&	32.82	&	5.80E-21	&		&	4.42E+08	&	6.5E+30	&	1.17E+02	&		&	28.76	&		&		&	26.96	\\
60	&	J2229+6114	&	51.62	&	7.83E-14	&	2.25	&	2.03E+12	&	2.2E+37	&	1.39E+05	&	33.12	&		&	34.29	&		&	33.40	\\
61	&	J2337+6151	&	495.37	&	1.93E-13	&	0.69	&	9.91E+12	&	6.3E+34	&	7.64E+02	&	31.46	&		&		&		&	30.35	\\
\hline
\end{supertabular}
\end{landscape}
\end{document}